\documentclass[prd,showkeys,floatfix,nofootinbib,
               preprint,12pt,
               tightenlines,fleqn]{revtex4} 

\usepackage{amsmath,amssymb,revsymb,graphicx,dcolumn}
\usepackage{leftidx}
\newcommand  {\version}{v3}        

\newcommand{\beq}{\begin{equation}}
\newcommand{\eeq}{\end{equation}}
\newcommand{\beqa}{\begin{eqnarray}}
\newcommand{\eeqa}{\end{eqnarray}}
\newcommand{\bsubeqs}{\begin{subequations}}
\newcommand{\esubeqs}{\end{subequations}}

\begin{document}
\noindent arXiv:1602.01475
\hfill KA--TP--04--2016\;(\version)\vspace*{8mm}\newline
\title[]
      {Low-energy electromagnetic radiation as \\ 
      an indirect probe of black-hole evaporation\vspace*{5mm}}
\author{Slava Emelyanov}
\email{viacheslav.emelyanov@kit.edu}
\affiliation{Institute for Theoretical Physics,\\
Karlsruhe Institute of Technology (KIT),\\
76131 Karlsruhe, Germany\\}

\begin{abstract}
\vspace*{2.5mm}\noindent
We study the influence of black-hole evaporation on light propagation. The framework employed
is based on the non-linear QED effective action at one-loop level. We show that the light-cone 
condition is modified for low-energy radiation due to black-hole evaporation. We discuss conditions 
under which the phase velocity of this low-energy radiation is greater than $c$. We also compute 
the modified light-deflection angle, which turns out to be significantly different from the standard GR 
value for black-hole masses in the range $M_\text{Pl} \ll M \lesssim 10^{19}\;M_\text{Pl}$.
\end{abstract}


\date{\today}

\maketitle

\section{Introduction}

A propagation of light in a non-trivial, i.e. non-Minkowskian, quantum state can be modified in quantum 
electrodynamics (QED). Moreover, the field operator $\hat{A}_\mu(x)$ of the electromagnetic field can 
have a non-standard structure due to non-trivial boundary conditions that leads to a modification of its 
action on the Minkowski vacuum. As a consequence, the photon propagator alters as well. Specifically, 
a low-energy electromagnetic wave propagating through a thermal gas turns out to be 
subluminal~\cite{Tarrach,Barton}, while superluminal when propagating in-between conducting plates 
in the Casimir set-up~\cite{Scharnhorst}. These two quantum effects can be described at the same time by 
considering the effective action of the electromagnetic field with integrated out fermion degrees of 
freedom~\cite{Barton}. It was further realised that a sign of the renormalized energy density and 
pressure determines whether the phase velocity of the low-energy electromagnetic radiation is greater or 
smaller than $c$ as measured under the standard conditions~\cite{Latorre&Pascual&Tarrach,Dittrich&Gies}.

In curved spacetime extra curvature-dependent terms appear in the effective action in quantum 
electrodynamics~\cite{Drummond&Hathrell}. At the leading $\alpha$-order these terms are 
quadratic with respect to the field strength of the electromagnetic field. This implies in particular that the 
Drummond-Hathrell term is oblivious to the quantum state at the $\alpha$-order approximation, 
but not to the spacetime geometry.

In the current paper we study the Maxwell field equations modified by the Euler-Heisenberg term as well as the 
Drummond-Hathrell term in the Schwarzschild black-hole 
geometry~\cite{Euler&Heisenberg,Drummond&Hathrell,Bastianelli&Davila&Schubert}. 
Under the assumption the vector-field operator modifies 
when the black-hole horizon forms, rather than the Fock space representation of the field operators 
as argued in~\cite{Emelyanov}, one can \emph{a priori} expect 
a non-trivial effect in spacetime with the black hole analogous to that in-between the conducting plates. Thus, our 
main concern in this paper is to investigate how quantum fluctuations of the electromagnetic 
field in the form of 
the Hawking radiation ($\langle \hat{T}_{\mu\nu}\rangle \neq 0$) influence the propagation of the long-wavelength radiation in quantum electrodynamics.

Throughout this paper the fundamental constants are set to unity, $c = G = k_\text{B} = \hbar = 1$.

\section{Effective field equations}
\label{sec:effective action}

Integrating out fermion degrees of freedom in QED, we obtain a non-linear
effective action for the electromagnetic field alone. This is exactly what we mean by the non-linear QED. 
In curved spacetime this leads to adding
the Drummond-Hathrell term~\cite{Drummond&Hathrell} and the
Euler-Heisenberg term~\cite{Euler&Heisenberg}
to the standard Maxwell action (see also~\cite{Bastianelli&Davila&Schubert}). 
We denote this effective action as $\Gamma_\text{eff}[A,g]$ below. The vector-field 
equation is thus modified in quantum electrodynamics and reads
\beqa\label{eq:mfe-classical}
\nabla_{\mu}F^{\mu\nu} - 4\left(\frac{4\alpha^2}{90m_e^4}\,F^{\mu\nu}F_{\lambda\rho}
+ \frac{7\alpha^2}{90m_e^4}\,\tilde{F}^{\mu\nu}\tilde{F}_{\lambda\rho} - 
\frac{\alpha}{360\pi m_e^2}\,R_{\;\;\;\;\lambda\rho}^{\mu\nu}\right)\nabla_{\mu}F^{\lambda\rho} &=& 0\,,
\eeqa
where $\alpha$ is the fine structure constant, $m_e$ the electron mass.
We have taken into account that $R_{\mu\nu} = 0$ in the Schwarzschild geometry 
\beqa\label{eq:sch}
ds^2 &=& f(r)dt^2 - \frac{dr^2}{f(r)} - r^2d\Omega^2\,, \quad f(r) \;=\; 1 - 2M/r\,,
\eeqa
and neglected higher-order derivative terms of the electromagnetic field strength $F_{\mu\nu}$ in order to have 
the same order of the approximation in the Euler-Heisenberg and the Drummond-Hathrell action.
We have also omitted the derivative of the Riemann tensor focusing only on the light wavelengths 
$\lambda_\gamma$ being much smaller than a characteristic curvature scale $\lambda_c$. 
Furthermore, the Euler-Heisenberg action is valid for the light wavelengths being much larger than 
the Compton length of the electron $\lambda_e$. Thus, the equation~\eqref{eq:mfe-classical} must be reliable 
in the regime $\lambda_e \ll \lambda_\gamma \ll \lambda_c$.

The vector-field equation \eqref{eq:mfe-classical} follows from variation of $\Gamma_\text{eff}[A,g]$ with 
respect to $A_{\mu}(x)$ and, thus, is classical in the sense that $A_\mu(x)$ is not quantised. 
However, the full effective action $\Gamma_\text{1PI}[A,g]$ differs from
$\Gamma_\text{eff}[A,g]$. We now want to take into account the influence of quantum fluctuations of the field $A_{\mu}(x)$ on the light propagation. Since we do not know an exact expression of $\Gamma_\text{1PI}[A,g]$, 
we follow~\cite{Mukhanov} to compute one-loop correction to the classical non-linear equation. Specifically, we
consider $\hat{A}_{\mu}'(x) = A_{\mu}(x) + \hat{a}_{\mu}(x)$, such that $\hat{a}_{\mu}(x)$ has an ordinary photon propagator in the Schwarzschild geometry. Substituting $\hat{A}_{\mu}'(x)$ in \eqref{eq:mfe-classical} and taking then its vacuum expectation value we get at the linear order in $A_{\mu}(x)$ in the one-loop approximation
\beqa\label{eq:mfe-quantum}
\nabla_{\mu}F^{\mu\nu} - 4\left(\frac{4\alpha^2}{90m_e^4}\,\langle \hat{f}^{\mu\nu}\hat{f}_{\lambda\rho}\rangle_\text{ren}
+ \frac{7\alpha^2}{90m_e^4}\,\langle\hat{\tilde{f}}^{\mu\nu}\hat{\tilde{f}}_{\lambda\rho}\rangle_\text{ren} - 
\frac{\alpha}{360\pi m_e^2}\,R_{\;\;\;\;\lambda\rho}^{\mu\nu}\right)\nabla_{\mu}F^{\lambda\rho}
&=& 0\,,
\eeqa
where we have neglected terms being of the order of $\lambda_e/\lambda_c \ll 1$ and 
$\lambda_\gamma/\lambda_c \ll 1$. It is worth mentioning  that this procedure of deriving~\eqref{eq:mfe-quantum} 
is equivalent to the background-field method of taking into account quantum field fluctuations at one-loop 
level (e.g., see~\cite{Schwartz}).

A few remarks are in order. First, the role of the fermion field is to provide self-interacting terms for the field 
$A_\mu(x)$ in the effective action $\Gamma_\text{eff}[A,g]$. Second, the effect of our interest crucially depends 
on having quantum fluctuations of the electromagnetic field. Third, the~equation \eqref{eq:mfe-quantum} 
follows from the 1PI effective action computed at the lowest-order approximation by promoting the electromagnetic 
field to a quantum operator in $\Gamma_\text{eff}[A,g]$. Fourth, all non-linear terms with respect to $A_\mu(x)$ 
in~\eqref{eq:mfe-quantum} have been omitted, because we want to study how low-energy electromagnetic waves 
behave in the vacuum when propagating through spacetime. In other words, these waves are our probe or test of
the electromagnetic properties of the vacuum (e.g., see~\cite{Dittrich&Gies1}).

Now the simplest way to obtain the light-cone condition is to employ the geometric optics approximation. This yields 
\beqa\label{eq:polarization}
k^2\epsilon^{\nu} - 8\left(\frac{4\alpha^2}{90m_e^4}\,\langle \hat{f}^{\mu\nu}\hat{f}_{\lambda\rho}\rangle_\text{ren}
+ \frac{7\alpha^2}{90m_e^4}\,\langle\hat{\tilde{f}}^{\mu\nu}\hat{\tilde{f}}_{\lambda\rho}\rangle_\text{ren} - 
\frac{\alpha}{360\pi m_e^2}\,R_{\;\;\;\;\lambda\rho}^{\mu\nu}\right)
k_{\mu}k^{\lambda}\epsilon^{\rho} &=& 0\,,
\eeqa
where we have used the Bianchi identity $k_{(\mu}F_{\lambda\rho)} = 0$ with $k_{\mu}$ being a 
wave vector, i.e. $\nabla_{\mu}F^{\lambda\rho} = ik_{\mu}F^{\lambda\rho}$, and
the vector $\epsilon^{\mu}$ specifies light polarisation in the Lorentz gauge.

\subsection{Modified radial propagation}

A computation of the light-cone condition for the radial propagation of the electromagnetic wave is considerably simplified 
in the Newman-Penrose formalism. Accordingly, one introduces the null tetrad 
$e_a^{\mu} = \{l^{\mu},n^{\mu},m^{\mu},\bar{m}^{\mu}\}$, such that $l^{\mu}n_{\mu} = -m^{\mu}\bar{m}_{\mu} = 1$
and the rest possible products vanish. Thus, we choose 
\beqa
k^{\mu} &=& l^{\mu} + \delta{l}^{\mu}\, \quad \text{and} \quad
\epsilon^{\mu} \;=\; \alpha_1 m^{\mu} + \alpha_2 \bar{m}^{\mu}
\eeqa
for the radial propagation. Substituting these in~\eqref{eq:polarization} and looking for a non-trivial solution for coefficients $\alpha_1$ and $\alpha_2$, 
we find that there exist two non-trivial polarisations $\epsilon_\pm^{\mu}$, such those
\beqa\label{eq:polarization:radial}
k^2 + \frac{4\alpha^2}{45m_e^4}\,\epsilon_{\pm}\langle \hat{T}_{\mu\nu} \rangle k^{\mu}k^{\nu} &=& 0\,,
\eeqa
where $\langle \hat{T}_{\mu\nu} \rangle$ is a renormalized stress tensor for $\hat{a}_\mu(x)$ and
\beqa
\epsilon_+ &=& 4\, \quad \text{and} \quad \epsilon_- \;=\; 7\,.
\eeqa
It is worth noting that the Drummond-Hathrell action does not
influence the radial light propagation~\cite{Drummond&Hathrell}.
The higher-order curvature-dependent terms have 
also no influence on the radial light propagation for the Schwarzschild black holes~\cite{Shore02a,Shore02b}. 

The formula~\eqref{eq:polarization:radial}
can be employed to get a change of the phase velocity of the low-energy electromagnetic radiation 
due to non-trivial renormalized stress tensor $\langle \hat{T}_{\mu\nu} \rangle$ of the quantum field in the 
thermal state as well as in-between the conducting plates~\cite{Tarrach,Scharnhorst,Barton,Latorre&Pascual&Tarrach}.
\begin{figure}[t]
\includegraphics[width=8.0cm]{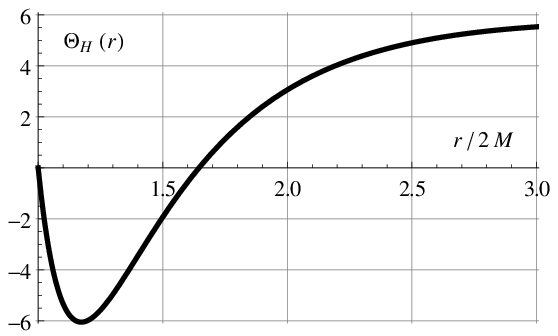}
\includegraphics[width=8.0cm]{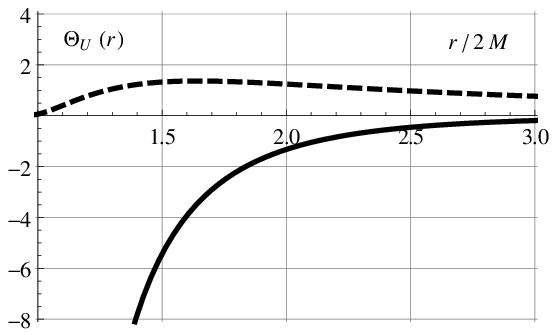}
\caption{Left: $\Theta_H(r) = (\langle \hat{T}_t^t \rangle_H - \langle \hat{T}_r^r \rangle_H)/\gamma$ as a function of 
$r$, where $\gamma = \frac{\pi^2}{45}T_H^4$, $T_H$ is the Hawking temperature~\cite{Jensen&Ottewill}. The
distance from the black-hole center when $\Theta_H(r)$ vanishes is approximately $r_n \approx 3.3{\times}M$. 
Right: $\Theta_U(r) = (\langle \hat{T}_t^t \rangle_U + 2\varepsilon f(r)\langle \hat{T}_r^t \rangle_U 
- \langle \hat{T}_r^r \rangle_U)/\gamma$ as a function of $r$~\cite{Jensen&McLaughlin&Ottewill,Matyjasek}. 
The solid and dashed line correspond to the outgoing and ingoing light wave for which $r_n = +\infty$ and 
$r_n = 0$, respectively.}
\end{figure}

We now apply that formula for the radially propagating electromagnetic wave in the background of the Hawking
radiation. We find
\beqa
\frac{\delta{c}}{c} &=& - \frac{2\alpha^2}{45m_e^4}\,\epsilon_{\pm}
\big(\langle \hat{T}_t^t \rangle + 2\varepsilon f(r)\langle \hat{T}_r^t \rangle - \langle \hat{T}_r^r \rangle\big)\,,
\eeqa
where $f(r)$ is a lapse function given in the equation~\eqref{eq:sch}, and $\varepsilon$ is either $+1$ or $-1$ for an
outgoing or ingoing light wave, respectively. 

In the case of the eternal Schwarzschild black hole physical vacuum corresponds to the 
Hartle-Hawking state which is regular on both past and future horizon. Employing  
results of~\cite{Jensen&Ottewill} for the renormalized stress tensor of the electromagnetic 
field in the Hartle-Hawking state, we find that the radially outgoing or ingoing radiation is superluminal for 
$r \in (2M,\,r_n)$, but subluminal for $r > r_n$, where $r_n \approx 3.3{\times}M$ (see fig.~1). 
Thus, the superluminal radial propagation between $2M$ and $r_n$ resembles
that in the Scharnhorst effect~\cite{Scharnhorst}, although the analogy is not complete (see below). 
It is worth noting that the violation of the null energy condition in our case is qualitatively similar for the case 
of the scalar field model conformally coupled to gravity~\cite{Visser96}.

In the case of a physical black hole, i.e. the black hole formed through the collapse of matter, physical
vacuum corresponds to the Unruh state. We employ an approximate analytic expression of the
renormalized stress tensor in the Unruh state~\cite{Jensen&McLaughlin&Ottewill,Matyjasek} to
analyse the influence of quantum fluctuations on the light propagation. The outgoing radiation 
turns out to be superluminal at any distance from the black hole, while the ingoing one is
subluminal right up to the horizon. Qualitatively the same picture of the violation of the null energy condition 
holds for the conformal scalar field model~\cite{Visser97}.

However, the Euler-Heisenberg action in the case of the Unruh state starts to dominate over the Maxwell action for distances roughly less than 
\beqa\label{eq:distance}
2M\left(10^3\,\frac{M_\text{Pl}^2}{Mm_e}\right)^4
\eeqa
from the black-hole horizon, where $M_\text{Pl}$ is the Planck mass. Therefore, \eqref{eq:distance} is negligibly 
small if the black-hole mass $M$ is sufficiently large, i.e.
\beqa
M &\gg& 10^3\,\frac{M_\text{Pl}^2}{m_e} \;\;\approx\;\; 10^{25}M_\text{Pl} \;\;\approx\;\; 10^{-13}M_{\odot}\,.
\eeqa
Thus, the approximation is reliable even close to the horizon for the black-hole masses being
much larger than $10^{-15}M_{\odot}$. Note that the violation 
of the weak gravity approximation may occur in the vicinity of the horizon. 
However, this is not the case for the radially propagating light
whenever its wavelength $\lambda_\gamma$ is much smaller than $\lambda_c$.

\subsection{Modified light deflection}

We now consider an electromagnetic wave propagating in the $\theta = \frac{\pi}{2}$ plane. Working in 
the notations of~\cite{Shore07}, we obtain the same formula~\eqref{eq:polarization:radial}, but now with
$k^{\mu}$ given by the solution of the geodesic equation depending on an impact parameter $d$ plus a 
correction of the $\alpha^2$-order. This is only possible in our context if the Drummond-Hathrell term is 
omitted. We are interested in the regime when the correction to the light deflection induced by this term 
is negligibly small, we thus study in the following a value of the light deflection being only due to the 
Euler-Heisenberg term.

Computing the deflection of light in the weak gravity limit, i.e. $r \gg 2M$, we find in the leading-order 
approximation
\beqa\label{eq:deflection-angle-h}
\Delta\phi_H &\approx&
\left(1 - \frac{\epsilon_\pm\,\alpha^2}{(720\pi)^2}\left(\frac{M_\text{Pl}^2}{Mm_\text{e}}\right)^4\right)
\Delta\phi_\text{GR}
\eeqa
for the Hartle-Hawking state, where $\Delta\phi_\text{GR} = \frac{4M}{r_0}$ with $r_0$ being the closest distance to the black hole. It is worth noting that the term in the parentheses of~\eqref{eq:deflection-angle-h} is of the order of
the deviation of the phase velocity of the light wave from $c$.

Repeating these computations for the physical black hole, i.e. in the Unruh state, we find
\beqa\label{eq:deflection-angle-u}
\Delta\phi_U &\approx& \Delta\phi_\text{GR} - 
\frac{\epsilon_\pm\,\alpha^2}{960}\,LM^2\left(\frac{M_\text{Pl}^2}{Mm_\text{e}}\right)^4
(\Delta\phi_\text{GR})^2
\eeqa
where $L$ is a luminosity equaling $L \approx 2.68{\times}10^{-6}\frac{4\pi}{M^2}$ for the
electromagnetic field~\cite{Elster}.\footnote{Note that there is a correction of the order $(\Delta\phi_\text{GR})^2$ 
due to general relativity only which we have omitted as being small in comparison with 
the term due to the Euler-Heisenberg action.} Expressing this correction to the angle of the light deflection 
through the change of the phase velocity of the electromagnetic wave at $r = r_0 \gg 2M$, we obtain
\beqa
\Delta\phi_U &\approx& \Delta\phi_\text{GR} - \frac{3\pi}{8}\left|\frac{\delta{c}_U}{c}\right|
\;\;\approx\;\;\left(1 - \epsilon_\pm\frac{2M}{r_0}\left(3.32{\times}10^{19}\,\frac{M_\text{Pl}}{M}\right)^4\right)
\Delta\phi_\text{GR}\,.
\eeqa

The Drummond-Hathrell contribution to the light deflection is negligibly small with respect to the 
Euler-Heisenberg one if
\beqa\label{eq:dh-less-eh}
2M\left(1.28{\times}10^{19}\,\frac{M_\text{Pl}}{M}\right)^{-2} &\ll& r_0\,.
\eeqa
Note that the higher-order curvature/derivative terms are also suppressed in comparison with the 
Euler-Heisenberg term. The angle of the light deflection could be of the order one or even larger 
with respect to the standard result of general relativity (GR) if
\beqa\label{eq:annulus}
2M\epsilon_\pm^{\frac{1}{2}}\left(6.75{\times}10^{19}\,\frac{M_\text{Pl}}{M}\right)^2 &\lesssim & r_0
\;\lesssim\; 2M\epsilon_\pm\left(3.32{\times}10^{19}\,\frac{M_\text{Pl}}{M}\right)^4,
\eeqa
where the lower bound is due to our assumption $|\delta{c}_U/c| \lesssim 1/10$. Therefore, we come
to a conclusion that the black-hole evaporation considerably influences the light propagation if the black-hole 
mass is sufficiently small, i.e.
\beqa
M &\lesssim& 10^{19} M_\text{Pl} \;\approx\; 10^{-19}M_{\odot}\,.
\eeqa
Note that the condition~\eqref{eq:dh-less-eh} as well as the weak gravity condition are then automatically satisfied
if the black-hole mass lies in this range. However, the semi-classical approximation is reliable if the black hole is not
too small, namely $M \gg M_\text{Pl}$ should be fulfilled. Thus, the above effects of the black-hole evaporation 
on the low-energy electromagnetic wave propagation are still trustable if the black-hole mass $M$ is much bigger than the Planck mass $M_\text{Pl}$,
so that $M_\text{Pl} \ll M \lesssim 10^{19}\,M_\text{Pl}$.
\begin{figure}[t]
\includegraphics[width=14.0cm]{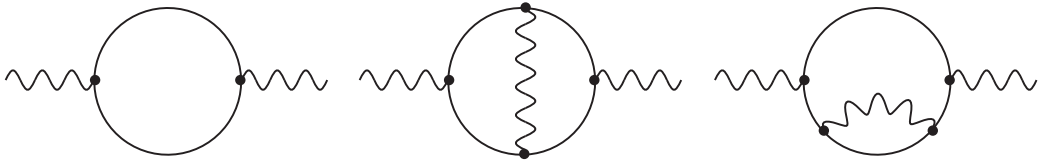}
\caption{Vacuum polarisation diagrams contributing to the photon self-energy up to the
$\alpha^2$-order.}
\end{figure}

\subsection{Two-loop dominance}
\label{sec:2ld}

In terms of the Feynman diagrams, the effect occurs due to the non-trivial modification of the photon propagator. 
The photon self-energy gets a correction at the level of two-loop diagrams depicted in fig.~2. This has 
been taken into account in the effective action. However, the electron/positron propagator also changes 
and, hence, makes a contribution to the modification of the photon propagator. 
This is the main difference in comparison with the Scharnhorst 
effect. The Hartle-Hawking state is the Kubo-Martin-Schwinger state~\cite{Haag} with respect to the Killing
vector $\partial_t$ for 
the field operators having a vanishing support in the causal complement to the ``right" Schwarzschild space. 
The renormalized stress tensor is mathematically indistinguishable, but not physically~\cite{Emelyanov}, from 
that for the thermal radiation at $T_H$ sufficiently far
from the black hole ($r \gg 2M$). Thus, the modification of the fermion propagator can be neglected as long as 
$M \gg 10^{21}M_\text{Pl}$,  i.e. when $T_H$ is much smaller than the electron mass $m_e$. However, the photon
acquires a thermal mass of the order of $\alpha^{\frac{1}{2}}T_H$ at $T_H \gg m_e$ due to the one-loop vacuum
polarisation diagram depicted in fig.~2. Therefore, our results for the Hartle-Hawking state are definitely reliable and 
does not lead to $m_\gamma \gg m_e$ if $T_H \ll m_e$ which is a regime of the two-loop dominance~\cite{Gies}.
\footnote{It is worth emphasising that eternal black holes are of no physical interests anyway, because these are not 
formed through the gravitational collapse. As one can see, if there is a small eternal black hole, then the effective 
photon mass is huge. This is physically unacceptable. Moreover, $\lambda_\gamma \lesssim \lambda_e$ should 
hold to have propagating light waves. This is beyond of our approximation.}
 
In the case of the Unruh state, the stress tensor vanishes as $T_H^4(2M/r)^2$ far from the hole. Taking this
into account as well as the structure of the one-loop vacuum polarisation diagram, the effective photon mass 
squared is expected to be of the order of $\alpha T_H^2(2M/r)^2$ far from the hole. At high temperatures, 
$T_H \gg m_e$, one thus has that our approximation is reliable if the light wavelength $\lambda_\gamma$ is much less than $\alpha^{\frac{1}{2}}\lambda_e (T_H/m_e)$, which is much greater than $\lambda_e$ for $M \lesssim 10^{19}\,M_\text{Pl}$. There are still propagating waves for
distances greater than the lower bound of $r_0$ given in~\eqref{eq:annulus}, because the effective plasma
frequence is suppressed by the geometrical factor $2M/r$. The $\alpha^3$-order term is expected to be of the order of $(T_H/m_e)^6(2M/r)^4$ which is suppressed far from the black hole. Specifically, the closest distance $r_0$ should approximately be larger than the lower bound given in~\eqref{eq:annulus}.

We analytically compute the effective photon mass $m_\gamma$ at one-loop level in QED in the
background of the small Schwarzschild black hole in~\cite{Emelyanov1}. 

\section{Concluding remarks}
\label{sec:concluding remarks}

We have analysed the influence of the vacuum polarisation induced by the black holes in 
quantum electrodynamics on the propagation of the low-energy electromagnetic radiation.
This results in the violation of the null energy condition and the superluminal/subluminal 
phase velocity of the radial outgoing/ingoing electromagnetic radiation, respectively. 

The black-hole evaporation might be observable through the angle of the light deflection. 
Specifically, for the black-hole masses in the range 
\beqa
M_\text{Pl} &\ll& M \;\lesssim\; 10^{19}\,M_\text{Pl}
\eeqa
and for the closest distance $r_0$ to the black hole lying in the annulus~\eqref{eq:annulus}, one can expect a significant deviation of the light-deflection angle from the standard GR value. Note that this angle is different for different types of the light polarisations. Practically, this implies that a source of unpolarised 
light has an image stretched in the direction from the hole. It is also worth emphasising this might be a physical 
effect being appropriate for discovering \emph{only} small black holes.\footnote{ Our estimate of the 
$\alpha^3$-order term made in Sec.~\ref{sec:2ld} may be too optimistic. If it turns out that the three-loop 
contribution to the light-cone condition is of the order of $(T_H/m_e)^6(2M/r)^2$, then the higher loop 
contributions are in general not negligible at $T_H \gg m_e$ as in the ordinary hot plasma. This would imply 
that the perturbation theory gets out of control. Thus, the perturbation series needs to be resummed. 
Nevertheless, the two-loop dominance still occurs for black-hole masses 
$10^{18}\,M_\text{Pl} \lesssim M \lesssim 10^{19}\,M_\text{Pl}$.}

The electromagnetic wave moving along a circular geodesic around the black hole propagates with the phase 
velocity less than $c$ at $r \gtrsim 27.9{\times}M$. The circular velocity approaches $c$ as $(2M/r)^2$
at $r \gg 2M$. The radial outgoing light velocity approaches $c$ as $(2M/r)^5$, while the ingoing 
one as $(2M/r)^2$. The ratio of the phase velocities of the radial ingoing wave and the circular wave is
\beqa\label{eq:anisotropy}
\frac{c_{in,\pm}}{c_{cir,\mp}} &\approx& 1- \left(2.68{\times}10^{19}\frac{M_\text{Pl}}{M}\right)^4
\left(\frac{2M}{r}\right)^2\big(4\epsilon_\pm - \epsilon_\mp\big)\,,
\eeqa
where the indices $+$ and $-$ refer to the light polarisations. For the supermassive black hole in the center 
of the Milky Way, we find an extremely small value $10^{-121}$ of the anisotropy of the phase velocities. This 
is much less than the anisotropy due to the Drummond-Hathrell term, i.e. $10^{-84}$, because the
constraint \eqref{eq:dh-less-eh} is not fulfilled. Note that the anisotropy due to the black-hole evaporation 
becomes larger whenever the black hole is lighter and closer to earth.

As pointed out above, the imprint of the black-hole evaporation on the light propagation is due to the 
modification of the vector-field and fermion operator when the event horizon forms. This is analogous 
to the Casimir effect, wherein the modification is however due to the boundary conditions satisfied 
by the electromagnetic field on the conducting plates (the fermion propagator is unaltered). This picture 
seams to be consistent with a unitary black-hole evaporation~\cite{Emelyanov}. 

The Drummond-Hathrell term leads to the violation of the strong principle of equivalence, whereas the 
Euler-Heisenberg term is completely consistent with this principle. These terms allow in particular to 
have a superluminal propagation of the low-energy electromagnetic radiation. However, the superluminality here 
does not necessary imply a violation of causality~\cite{Shore96,Shore02a,Liberati&Sonego&Visser,Shore06,Hollowood&Shore,Hollowood&Shore&Stanley}.
To establish whether causality is not broken, one needs to know how high-energy photons 
propagate in the background of the evaporating black hole. Indeed, the wave-front velocity corresponding
to the signal velocity is given by the phase velocity of the high-energy radiation~\cite{Leontovich}. 
This is certainly beyond of our approximation. Nevertheless, one 
might expect that causality is \emph{not} violated as we have started with a perfectly causal theory, namely QED, in 
the classical geometrical background.

It is not obvious whether the so-called ``horizon" theorem~\cite{Shore02a} still holds for the radially outgoing 
low-energy radiation, i.e. whether the light cone remains $k^2 = 0$ for the outgoing light wave on the event 
horizon of a
black hole formed through the gravitational collapse. Although the perturbation theory breaks down near 
$r = 2M$ for the 
outward light, one can still have $k^2 < 0$ for the outgoing radiation very close to the horizon for 
sufficiently large black holes, $M \gg 10^{25}M_\text{Pl}$. However, we expect that this theorem still 
holds on the horizon, because $k^2 > 0$ close to it inside the hole as it can be shown by performing an
analytic continuation of the Schwarzschild coordinates inside the black hole.

\section*{
ACKNOWLEDGMENTS}
It is a pleasure to thank Frans Klinkhamer for valuable discussions and many helpful comments on 
an early version of this paper. We are grateful to the anonymous referee for many fruitful questions/comments
which helped to improve the manuscript.

\begin{appendix}
\end{appendix}


\begin{thebibliography}{99}

\bibitem{Tarrach}
R. Tarrach,
\hspace*{0mm}``Thermal effects on the speed of light,''
Phys. Lett. B{\bf 133}, 259 (1983).

\bibitem{Barton}
G. Barton,
\hspace*{0mm}``Faster-than-c light between parallel mirrors. The Scharnhorst effect rederived,''
Phys. Lett. B{\bf 237}, 559 (1990).

\bibitem{Scharnhorst}
K. Scharnhorst,
\hspace*{0mm}``On propagation of light in the vacuum between plates,''
Phys. Lett. B{\bf 236}, 354 (1990).

\bibitem{Latorre&Pascual&Tarrach}
J.I. Latorre, P. Pascual, R. Tarrach,
\hspace*{0mm}``Speed of light in non-trivial vacua,''
Nucl. Phys. B{\bf 437}, 60 (1995); arXiv:hep-th/9408016.

\bibitem{Dittrich&Gies}
W. Dittrich, H. Gies,
\hspace*{0mm}``Light propagation in non-trivial QED vacua,''
Phys. Rev. D{\bf 58}, 025004 (1998); arXiv:hep-ph/9804375.

\bibitem{Drummond&Hathrell}
I.T. Drummond, S.J Hathrell,
\hspace*{0mm}``QED vacuum polarization in a background gravitational field and its effect on the
velocity of photons,''
Phys. Rev. D{\bf 22}, 343 (1980).

\bibitem{Euler&Heisenberg}
W. Heisenberg, H. Euler,
\hspace*{0mm}``Folgerungen aus der Diracsche Theorie des Positrons,''
Z. Phys. {\bf 98}, 714 (1936).

\bibitem{Bastianelli&Davila&Schubert}
F. Bastianelli, J.M. D\'{a}vila, Ch. Schubert,
\hspace*{0mm}``Gravitational corrections to the Euler-Heisenberg lagrangian,''
JHEP03, 086 (2009); arXiv:hep-th/0812.4849.

\bibitem{Emelyanov}
S. Emelyanov,
\hspace*{0mm}``Can gravitational collapse and black-hole evaporation be a unitary process after all?,''
arXiv:hep-th/1507.03025.

\bibitem{Mukhanov}
V.F. Mukhanov,
\hspace*{0mm}{\sl Physical Foundations of Cosmology}
(Cambridge University Press, 2005).

\bibitem{Schwartz}
M.D. Schwartz,
\hspace*{0mm}{\sl Quantum Field Theory and the Standard Model}
(Cambridge University Press, 2014).

\bibitem{Dittrich&Gies1}
W. Dittrich, H. Gies,
\hspace*{0mm}{\sl Probing the quantum vacuum. Perturbative effective action approach 
in Quantum Electrodynamics and its application} (Springer-Verlag, 2000).

\bibitem{Shore02a}
G.M. Shore,
\hspace*{0mm}``Faster than light photons in gravitational fields II. Dispersion and vacuum polarisation,''
Nucl. Phys. B{\bf 633}, 271 (2002); arXiv:gr-qc/0203034.

\bibitem{Shore02b}
G.M. Shore,
\hspace*{0mm}``A local effective action for photon-gravity interactions,''
Nucl. Phys. B{\bf 646}, 281 (2002); arXiv:gr-qc/0205042.

\bibitem{Jensen&Ottewill}
B.P. Jensen, A. Ottewill,
\hspace*{0mm}``Renormalized electromagnetic stress tensor in Schwarzschild spacetime,''
Phys. Rev. D{\bf 39}, 1130 (1989).

\bibitem{Visser96}
M. Visser,
\hspace*{0mm}``Gravitational vacuum polarization. I. Energy conditions in the Hartle-Hawking vacuum,''
Phys. Rev. D{\bf 54}, 5103 (1996); arXiv:gr-qc/9604007.

\bibitem{Jensen&McLaughlin&Ottewill}
B.P. Jensen, J.G. McLaughlin, A.C. Ottewill,
\hspace*{0mm}``Renormalized electromagnetic stress tensor for an evaporating black hole,''
Phys. Rev. D{\bf 43}, 4142 (1991).

\bibitem{Matyjasek}
J. Matyjasek,
\hspace*{0mm}``Stress-energy tensor of an electromagnetic field in Schwarzschild spacetime,''
Phys. Rev. D{\bf 55}, 809 (1997),
\hspace*{0mm}``$\langle T_{\nu}^{\mu}\rangle_\text{ren}$ of the quantized conformal fields
in the Unruh state in Schwarzschild spacetime,''
Phys. Rev. D{\bf 59}, 044002 (1999); arXiv:gr-qc/9808019.

\bibitem{Visser97}
M. Visser,
\hspace*{0mm}``Gravitational vacuum polarization. IV. Energy conditions in the Unruh vacuum,''
Phys. Rev. D{\bf 56}, 936 (1997); arXiv:gr-qc/9703001.

\bibitem{Liberati&Sonego&Visser}
S. Liberati, S. Sonego, M. Visser,
\hspace*{0mm}``Faster than c signals, special relativity, and causality,''
Ann. Phys. {\bf 298}, 167 (2002); arXiv:gr-qc/0107091.

\bibitem{Shore06}
G.M. Shore,
\hspace*{0mm} Causality and Superluminal Light, in
\emph{Time and Matter},
edited by I.I. Bigi, M. Faessler (World Scientific Publishing, 2006).

\bibitem{Shore07}
G.M. Shore,
\hspace*{0mm}``Superluminality and UV completion,''
Nucl. Phys. B{\bf 778}, 219 (2007); arXiv:hep-th/0701185.

\bibitem{Elster}
T. Elster,
\hspace*{0mm}``Vacuum polarization near a black hole creating particles,''
Phys. Lett. A{\bf 94}, 205 (1982).

\bibitem{Haag}
R. Haag,
\hspace*{0mm}{\sl Local quantum physics. Fields, Particles, Algebras}
(Springer-Verlag, 1996).

\bibitem{Gies}
H. Gies,
\hspace*{0mm}``QED effective action at finite temperature: Two-loop dominance,''
Phys. Rev. D{\bf 61}, 085021 (2000); arXiv:hep-ph/9909500.

\bibitem{Emelyanov1}
S. Emelyanov,
\hspace*{0mm}``Effective photon mass from black-hole formation,''~arXiv:hep-th/1603.01148.

\bibitem{Shore96}
G.M. Shore,
\hspace*{0mm}``Faster than light photons in gravitational fields - Causality, anomalies and horizons,''
Nucl. Phys. B{\bf 460}, 379 (1996); arXiv:gr-qc/9504041.

\bibitem{Hollowood&Shore}
T.J. Hollowood, G.M. Shore,
\hspace*{0mm}``Causality and micro-causality in curved spacetime,''
Phys. Lett. B{\bf 655}, 67 (2007); arXiv:hep-th/0707.2302,
\hspace*{0mm}``The refractive index of curved spacetime: The fate of causality in QED,''
Nucl. Phys. B{\bf 795}, 138 (2008); arXiv:hep-th/0707.2303,
\hspace*{0mm}``The causal structure of QED in curved spacetime: Analyticity and the refractive index,''
JHEP12, 091 (2008); arXiv:hep-th/0806.1019.

\bibitem{Hollowood&Shore&Stanley}
T.J. Hollowood, G.M. Shore, R.J. Stanley,
\hspace*{0mm}``The refractive index of curved spacetime II: QED, Penrose limits and black holes,''
JHEP08, 089 (2009); arXiv:hep-th/0905.0771.

\bibitem{Leontovich}
M.A. Leontovich,
\hspace*{0mm} Wave-front propagation, in
\emph{Lectures on Optics, Relativity, and Quantum Mechanics},
edited by L.I. Mandelshtam (Nauka, Moscow, 1972) (in Russian).

\end{thebibliography}
\end{document}